# Nanometer-scale lateral p-n junctions in graphene/α-RuCl₃ heterostructures

Daniel J. Rizzo[1,†], Sara Shabani[1,†], Bjarke S. Jessen[1,2,†], Jin Zhang[3,†], Alexander S. McLeod[1,#], Carmen Rubio-Verdú[1], Francesco L. Ruta[1,4], Matthew Cothrine[5], Jiaqiang Yan[5,6], David G. Mandrus[5,6], Stephen E. Nagler[7], Angel Rubio[3,8,9,*], James C. Hone[3], Cory R. Dean[1], Abhay N. Pasupathy[1,10*], D.N. Basov[1,*]

[1]Department of Physics, Columbia University, New York, NY, 10027, USA

[2]Department of Mechanical Engineering, Columbia University, New York, NY, 10027, USA

[3]Theory Department, Max Planck Institute for Structure and Dynamics of Matter and Center for Free-Electron Laser Science, 22761 Hamburg, Germany

[4]Department of Applied Physics and Applied Mathematics, Columbia University, New York, NY, 10027, USA

[5]Department of Materials Science and Engineering, University of Tennessee, Knoxville, Tennessee 37996, USA

[6]Materials Science and Technology Division, Oak Ridge National Laboratory, Oak Ridge, Tennessee 37831, USA

[7]Neutron Scattering Division, Oak Ridge National Laboratory, Oak Ridge, Tennessee 37831, United States

[8]Center for Computational Quantum Physics, Flatiron Institute, New York, New York 10010, USA

[9]Nano-Bio Spectroscopy Group, Universidad del País Vasco UPV/EHU, San Sebastián 20018, Spain

[10]Condensed Matter Physics and Materials Science Department, Brookhaven National Laboratory, Upton, NY 11973, USA

[†]Contributed equally

[#]Present Address: School of Physics and Astronomy, University of Minnesota, Minneapolis, MN 55455

*Correspondence to: db3056@columbia.edu, apn2108@columbia.edu, and angel.rubio@mpsd.mpg.de

**Abstract**

The ability to create high-quality lateral p-n junctions at nanometer length scales is essential for the next generation of two-dimensional (2D) electronic and plasmonic devices. Using a charge-transfer heterostructure consisting of graphene on α-RuCl₃, we conduct a proof-of-concept study demonstrating the existence of intrinsic nanoscale lateral p-n junctions in the vicinity of graphene nanobubbles. Our multi-pronged experimental approach incorporates scanning tunneling microscopy (STM) and spectroscopy (STS) and scattering-type scanning near-field optical microscopy ($s$-SNOM) in order to simultaneously probe both the electronic and optical responses of nanobubble p-n junctions. Our STM and STS results reveal that p-n junctions with a band offset of more than 0.6 eV can be achieved over lateral length scale of less than 3 nm, giving rise to a staggering effective in-plane field in excess of $10^8$ V/m. Concurrent $s$-SNOM measurements confirm the utility of these nano-junctions in plasmonically-active media, and validate the use of a point-scatterer formalism for modeling surface plasmon polaritons (SPPs). Model *ab initio* density functional theory (DFT) calculations corroborate our experimental data and reveal a combination of sub-angstrom and few-angstrom decay processes dictating the dependence of charge transfer on layer separation. Our study provides experimental and conceptual foundations for the use of charge-transfer interfaces such as graphene/α-RuCl₃ to generate p-n nano-junctions.

**Introduction**

Nanoscale lateral p-n junctions in graphene present promising routes for investigating fundamental quantum phenomena such as Andreev reflection[1,2], whispering gallery mode resonators[3,4], quantum dots[5-9], Veselago lensing[10,11] and photonic crystals[12]. The ability to realize nanoarchitectures capable of hosting these properties relies on precise control over the lateral p-n junction size – ideally down to atomic length scales. Despite the potential advantages of tailored nanometer junctions, attempts to realize sharp and clean interfacial junctions in graphene-based devices have been limited to > 20 nm[11,13] and lack the nominal potential profile for yielding high-quality devices. Conventional techniques such as local back gating[14], ion implantation[15,16], and adatoms[17] are practically challenging to implement and can be accompanied by an increase in disorder, reduction in mobility, and surface contamination. Moreover, the maximum charge carrier density achievable with these approaches is typically limited to less than $5 \times 10^{12}$ cm$^{-2}$,[18,19] restricting the potential gradients accessible with these techniques.

Recent theoretical[20,21] and experimental[22-25] work on graphene/$\alpha$-RuCl$_3$ heterostructures demonstrates that the Dirac-point energy ($E_{\text{Dirac}}$) in graphene will experience a massive shift (~0.6 eV) due to work function-mediated interlayer charge transfer with the underlying $\alpha$-RuCl$_3$. While transport measurements suggest a high degree of interlayer charge transfer[23] in graphene/$\alpha$-RuCl$_3$ heterostructures (>$10^{13}$ cm$^{-2}$), they have not revealed the lateral dimensions of this charging process. On the other hand, analysis of the plasmonic behavior of graphene/$\alpha$-RuCl$_3$ in the vicinity of nanobubbles suggests that boundaries between highly doped and pristine graphene are no wider than 50 nm[22]. Raman maps conducted on these heterostructures produce similar constraints on the maximum size of lateral charge modulation boundaries[24]. However, a detailed understanding of the nanoscale spatial dependence of interlayer charge transfer between graphene and α-RuCl$_3$ necessitates use of a high-resolution local probe.

In order to elucidate the intrinsic lateral and vertical length scales associated with interlayer charge transfer in graphene/α-RuCl$_3$ heterostructures, we employ two complementary imaging and spectroscopic techniques: scanning tunnelling microscopy and spectroscopy (STM/STS) and scattering-type scanning near-field optical microscopy (s-SNOM). STM and STS are ideal probes for studying lateral junction interfaces (e.g. p-n , p-p', p-i-p, etc.) with atomic resolution and provide information about the local electronic structure (in particular, $E_{Dirac}$ in graphene). On the other hand, s-SNOM uses hybrid light-matter modes known as surface plasmon polaritons (SPPs) to probe the local conductivity in graphene. This multi-messenger experimental approach provides a multifaceted view of the fundamental length scales associated with interlayer charge transfer as encoded in both the electronic and plasmonic responses of graphene/α-RuCl$_3$.

We use nanobubbles that arise spontaneously at the graphene/α-RuCl$_3$ heterostructure interface during fabrication as a testbed for probing the in-plane and out-of-plane behavior of interlayer charge transfer. Differential conductivity (d$I$/d$V$) maps and point spectroscopy performed at the boundary of nanobubbles reveal that highly p-doped and intrinsically n-doped graphene are separated by a lateral distance of ~3 nm and vertically by ~0.5 nm, generating internal fields on the order of $10^8$ V/m that are largely confined to the graphene plane. At the same time, the rapid change in the graphene conductivity in the vicinity of nanobubbles acts as a hard plasmonic barrier that reflects SPPs generated during s-SNOM measurements, as observed previously[22]. The results of STS measurements inform our interpretation of the s-SNOM data and permit us to further develop our model for the complex-valued near-field signal in the vicinity of nanobubbles using a perturbative point-scatterer approach. Our results are well supported by first-principles density-functional theory (DFT) calculations, which reveal the origin of the sharp spatial profile of interlayer charge transfer at the boundary of nanobubbles.

**Results and Discussion**

The graphene/α-RuCl₃ heterostructures studied herein were fabricated using dry transfer techniques from components isolated using exfoliation from single-crystal sources (see methods and Fig. S1 for a detailed description of the fabrication process). The resulting heterostructure consists of large regions of graphene forming a flat interface with the underlying α-RuCl₃, which are occasionally interrupted by graphene nanobubbles (Fig. 1A) (see Fig. S2 for STM topographic overview).

A high magnification topographic STM image of a characteristic graphene nanobubble is shown in Fig. 1B. As observed with STM topography, the typical heights of nanobubbles studied in this work were between 1 to 3 nm, while the radius ranged from 20 to 80 nm. Topographic images collected with an atomic force microscope (AFM) used during *s*-SNOM measurements yield similar nanobubble dimensions (Fig. S2). On the other hand, near-field images of these same nanobubbles collected using *s*-SNOM reveal larger circular features that extend over lateral distances of several hundred nanometers (Fig. 1C). The oscillatory nature of the near-field signal moving radially from nanobubbles is consistent with the presence of SPPs that are either being launched or reflected from these locations, giving rise to modulations in the near-field signal that extend far beyond the nanobubble area. It has been suggested that these plasmonic features arise due to discontinuities in the graphene conductivity associated with local modulation of charge carrier density[22], though the precise nature of this profile demands further scrutiny with STM and STS.

In order to gain insight into the spatial dependence of interlayer charge transfer, we performed a series of STM and STS measurements in the vicinity of four different graphene nanobubbles (Figs. 2, S3). Figure 2A shows two representative point spectra collected on a flat interface of graphene/$\alpha$-RuCl$_3$ (red) and on a nanobubble (blue) as indicated by the crosshairs in the topographic inset. The spectrum taken on the nanobubble (blue) is characteristic of slightly intrinsically n-doped graphene since the Dirac point is located at –0.1 eV relative to the Fermi energy $E_F$. This spectrum acts as a reference point for the pristine graphene density of states. On the other hand, the d$I$/d$V$ intensity on the flat graphene/$\alpha$-RuCl$_3$ region (red) away from the nanobubble junction shows a shift in the Dirac point energy of $\Delta E_{Dirac}$ = +0.625 eV relative to pristine graphene suspended in the nanobubble. This massive shift in $E_{Dirac}$ corresponds to a hole density in graphene greater than $10^{13}$ cm$^{-2}$ resulting from interlayer charge transfer with $\alpha$-RuCl$_3$. We attribute the local minimum close to $E_F$ observed for both spectra to the ubiquitous inelastic tunneling gap that arises due to phonon mediated processes independent of the graphene charge carrier density[19]. This direct observation of heavily p-doped graphene on $\alpha$-RuCl$_3$ by STM is consistent with the previous optical and transport studies[22-25], and demonstrates that p-n junctions are formed at the boundaries of nanobubbles.

To visualize nanobubble p-n junctions, d$I$/d$V$ maps were conducted at biases corresponding to $E_{Dirac}$ for both the nanobubble and flat interface regions (Fig 2B). The spectroscopic map conducted at -100 mV associated with $E_{Dirac}$ of the nanobubble shows a high LDOS on the surrounding graphene/$\alpha$-RuCl$_3$ compared to the nanobubble area with a jump in the LDOS at the boundary between these two regions that occurs over a lateral length scale of approximately 3 nm (Fig. 2C). This is consistent with the expectation that the nanobubble should have a suppressed LDOS at its $E_{Dirac}$ compared to the surrounding highly doped regions. On the other hand, at +525 meV (i.e. $E_{Dirac}$ of the flat graphene/α-RuCl$_3$ interface) the LDOS is enhanced on the nanobubble compared to the surrounding flat graphene/α-RuCl$_3$ region due to the corresponding Dirac point minimum of the latter. A similarly abrupt shift in the LDOS at the nanobubble edge is observed at this energy (Fig. 2C). This behavior is characteristic of a nanometer-scale p-n interface in graphene located at the nanobubble boundary.

We then extracted the potential profile across the p-n junction and evaluated its sharpness. A representative d$I$/d$V$ line cut is shown in Fig. 2D and follows the cyan trajectory highlighted in the inset of Fig. 2A. Fig. 2D clearly shows that the local minimum of the Dirac point shifts abruptly at the boundary of the nanobubble from +0.525 eV to –0.1 eV over a length scale of only a few nanometers. To provide information about the correspondence between STM topography and the shift in $E_{Dirac}$, we compare the nanobubble topographic cross-section (denoted with a cyan dotted line in Fig. 2D) with the Dirac energy position (green dashed line). It is evident that the change in the graphene doping level occurs much more abruptly than the height profile of the nanobubble, implying that interlayer charge transfer is rapidly suppressed with interlayer separation. The lateral junction width is measured to be ~3 nm as indicated by the partially transparent blue and red rectangle in Fig. 2D. The lateral width of this depletion region is roughly one order of magnitude smaller than previously reported results on state-of-the-art split back gate devices[13]. To provide a step-by-step view of the evolution of $E_{Dirac}$ across the junction, a few spectra from the highlighted region of the line cut are shown in Fig. 2E. These individual spectra allow us to understand how the doping level develops over the 3 nm transition at the nanobubble boundary. Once the interface of the nanobubble is reached and the graphene begins to separate from the underlying $\alpha$-RuCl$_3$ layer, the minimum corresponding to the Dirac point at +0.525 eV rapidly shifts to lower biases. Beyond this point, $E_{Dirac}$ shifts more gradually until it reaches its minimum value of –100 mV. The dependence of the shift in $E_{Dirac}$ on the nanobubble height is shown explicitly in Fig. 4D, hinting that two distinct mechanisms govern the interlayer charge transfer process, giving rise to two characteristic vertical length scales.

Armed with the results of STM and STS experiments, we now return to *s*-SNOM images conducted on graphene nanobubbles. Data were collected on five different nanobubbles over a frequency range of 930 – 2280 cm$^{-1}$ (Fig. 3). Characteristic images of the near-field amplitude and phase for $\omega = 990$ cm$^{-1}$ are shown in Fig. 3A along with the associated nanobubble dimensions. Immediately outside the radius of the nanobubble, radial oscillations of both near-field channels decay as a function of distance as shown explicitly in the linecuts in Fig. 3C. As expected[22], the spacing between fringes clearly disperses with frequency (Fig. S4). In principle, these fringes could arise from SPPs generated on and propagating away from nanobubbles (so-called $\lambda_p$ fringes), from SPPs generated at the AFM tip that reflect from the nanobubble boundary ($\lambda_p/2$ fringes), or from both. Previous work on similar heterostructures would suggest the near-field behavior is primarily dominated by the latter[22].

To definitively resolve this question, it is useful to consider that the STS data provides unambiguous evidence that the entirety of the graphene nanobubble consists of nominally undoped graphene surrounded by highly-doped graphene with a boundary width on the order of only a few nanometers. We therefore model the *s*-SNOM data of a graphene nanobubble as a raster-scanned dipole over a circular conductivity depletion region surrounded by a bulk possessing high conductivity in a manner similar to our previous study[22] (Fig. 3B, see supplementary discussion for detailed model description). Expanding on this previous work, we now consider that the SPPs generated at the AFM tip during *s*-SNOM measurements may possess a wide range of wavelengths relative to the size of the nanobubble. At one extreme, the SPP wavelength is much larger than the nanobubble and can pass through with little to no scattering. Here, a maximum in both the near-field amplitude and phase is observed at the location immediately outside the nanobubble boundary. At the other extreme, the SPP wavelength is too small to effectively couple to a finite-sized tip, suppressing the generation of SPPs in the first place. At intermediate length scales where the SPP wavelength is on the order of several times the nanobubble dimensions, plasmonic reflections are observed that result in $\lambda_p/2$ fringes whose amplitude scale as $\left(\frac{r_{bubble}}{\lambda_p}\right)^2$, where $r_{bubble}$ is the nanobubble radius (Fig. 3B). In contrast to the behavior at large $\lambda_p$, here the near-field amplitude possesses a minimum corresponding to the region immediately surround the defect, while the phase has a maximum in this same location. A comparison of the experimental and simulated near-field images shown in Figs. 3A and B suggests that our experiment takes place in this intermediate regime where plasmonic reflections give rise to $\lambda_p/2$ fringes and the near-field amplitude has a minimum while the phase has a maximum in the region just outside the nanobubble (indicated by the black dashed boxed region in Fig. 3B). In principle, $\lambda_p$ fringes could exist concurrently as a result of light scattering directly from vacuum into the graphene from the nanobubble itself. Such fringes would have a systematic angular dependent amplitude enforced by the angle of the incident light projected onto the 2D plane. Since a systematic angular dependence is neither observed in near-field amplitude nor phase (Fig. S4), we rule out the possibility that $\lambda_p$ fringes are substantially contributing to the observed SPP oscillations.

An approximate representation of the radial dependence of the near-field amplitude can be derived by perturbatively treating the nanobubble as a point scatterer (i.e. a point at which interlayer charge transfer does not take place). This is a 2D analogue of Rayleigh scattering and may be useful for analysis of SPP dispersions in a manner analogous to quasiparticle interference (QPI) of 2D electronic states[26,27]. Within this framework, the scattered polariton field is used as a proxy for the near-field signal and has the functional form of $-A\left[H_1^{(1)}(\boldsymbol{q}_p r)\right]^2$, (here, $H_1^{(1)}$ is the Hankel function of the first kind of order one, $\boldsymbol{q}_p = q_1 + iq_2$ is the complex SPP wavevector, $r$ is the radial coordinate and $\boldsymbol{A}$ is a complex scaling factor) (see supplementary discussion for full derivation). The real and imaginary components of this function are simultaneously fit to the near-field amplitude and phase, respectively, using $\boldsymbol{A}$ and $\boldsymbol{q}_p$ as fitting parameters. The resulting model line profiles faithfully reproduce the experimental data (Fig. 3C). Repeating this fitting procedure for all experimental frequencies $\omega$ and all five bubbles yields the SPP dispersion $\omega(q_1)$ (Fig. 3D). The shape of the experimental dispersion is consistent with SPPs propagating in highly doped graphene.

Both experimental STM/STS and *s*-SNOM data provide corroborating evidence that interlayer charge transfer between graphene and α-RuCl₃ is eliminated in nanobubbles as a result of < 1 nm of interlayer separation. We now inquire into the precise mechanism by which this charge transfer takes place and how it is suppressed in nanobubbles through a series of DFT calculations on model graphene/α-RuCl₃ heterostructures. Specifically, we explored the role of an intermediate vacuum region between the two layers varied from 0 to 5 Å above the equilibrium separation (Fig. 4A). As reported previously[22], the shift in $E_{\text{Dirac}}$ for the graphene/α-RuCl₃ heterostructure with an equilibrium interlayer separation ($h_{\min} = 3.3$ Å) is observed to be 0.54 eV, in good agreement with the experimental data on flat interface regions (Fig. 4B). However, the theoretical shift in $E_{\text{Dirac}}$ effectively disappears once a vacuum spacer layer of just $\Delta h = h - h_{\min} = 5$ Å is introduced (Fig. 4B), revealing a rapid decay in the interlayer charge transfer as a function of layer separation. At intermediate layer separations, the theoretical dependence of $\Delta E_{Dirac}$ on the interlayer separation shows a rapid jump for $\Delta h < 1$ Å followed by a more gradual decay in the interlayer charge transfer at larger separations (Fig. 4C). The experimental counterpart to this data can be extracted from Fig. 2D to visualize $\Delta E_{Dirac}$, (and thus the magnitude of the interlayer charge transfer) as a function of the interlayer separation between graphene and α-RuCl$_3$. Here, the shift of the Dirac point energy, $\Delta E_{Dirac}$, is obtained from the local minima of each d$I$/d$V$ spectrum taken at a given height relative to the flat graphene/α-RuCl$_3$ region across the p-n junction. $\Delta E_{Dirac}$ is plotted as a function of height to quantify the effect of interlayer separation on doping level. Figure 4C demonstrates that the behavior of the model DFT calculation mirrors the experimental STS: both show two characteristic decay lengths of less than and on the order of a few angstroms, respectively. We speculate that the emergence of two characteristic length scales associated with interlayer charge transfer arises due to a dual mechanism associated with short-range interlayer tunneling and a long-range polarization effect between the layers.

The agreement between theory and experiment also shows that the magnitude of interlayer charge transfer is ostensibly agnostic to the surrounding in-plane charge environment (i.e., purely dependent on the layer separation). Thus, it would appear that there is little to no charge redistribution in the graphene plane across the nanobubble interface despite large differences in the local charge carrier density. To understand this, we return to the DFT calculations of model heterostructures with variable vacuum spacer layers and plot $\Delta E_{Dirac}$ relative to the vacuum energy (green curve in Fig. 4D). From this, it is clear that an electrostatic barrier comparable to the offset in $E_{Dirac} \sim 0.6$ eV emerges between the pristine nanobubble and the highly doped graphene/α-RuCl$_3$ region. Ultimately, this large electrostatic barrier enforces the sharp p-n junctions intrinsically generated in nanobubbles found in graphene/α-RuCl$_3$ heterostructures.

**Conclusion**

We have measured the electronic and photonic behavior of nanobubbles in graphene/α-RuCl$_3$ heterostructures, revealing massive shifts in the local interlayer charge transfer over lateral length scales of only a few nanometers. Such narrow p-n junctions in graphene have previously been inaccessible using standard doping techniques and have many potential applications for studying fundamental electronic structure properties in graphene and related materials. At the same time, our results demonstrate that work function mediated charge transfer is a viable route toward creating nanoscale conductivity features in graphene that actively influence the local plasmonic behavior at sub-wavelength length scales. The insights gained in our DFT calculations provide a detailed understanding of the dependence of charge transfer on interfacial separation, and reveal abrupt electrostatic barriers at nanobubble boundaries giving rise to nanometer-scale p-n junctions. This work provides the experimental and conceptual foundation for future device design, and validates the use of interstitial layers in charge-transfer heterostructures to predictively influence the local electronic and plasmonic behavior.

**Methods**

***Material Growth***: α-RuCl$_3$ crystals were grown by the sublimation of RuCl$_3$ powder sealed in a quartz tube under vacuum. About 1 g of powder was loaded in a quartz tube of 19 mm in outer diameter, 1.5 mm thick, and 10 cm long. The growth was performed in a box furnace. After dwelling at 1060 °C for 6 h, the furnace was cooled to 800 °C at a rate of 4 °C/h. Magnetic and specific heat measurements confirmed that the as-grown pristine crystal orders antiferromagnetically around 7 K. For more information, see ref. [28].

***Device Fabrication***: α-RuCl$_3$ is notoriously difficult to pick up using standard dry stacking techniques. To overcome this limitation, we modify the usual dry stacking procedure in the following ways: When exfoliating α-RuCl$_3$ onto SiO2, we avoid any plasma treatment of the SiO$_2$ prior to exfoliation. This reduces the adhesion of the α-RuCl$_3$ to the SiO$_2$ (albeit at the expense of the yield of large-area crystals, which were not needed in this experiment).

To pick up the α-RuCl$_3$, we employ PDMS stamps coated with polycarbonate (PC). The PC is heated above the glass-transition temperature ($T$g ~ 150 ℃) to 170 ℃, leaving the film in a low viscosity state. We then slowly cover the target α-RuCl$_3$ flake and leave the PC in contact with the α-RuCl$_3$ for at least 10 minutes to ensure high coverage. Next, we lower the temperature to below $T$g, solidifying the PC film around the α-RuCl$_3$ crystal and significantly increasing the chance of a successful pick-up. We note that the temperature should not be raised higher than the values provided here, as the α-RuCl$_3$ will readily decompose in ambient at temperatures above 200 ℃. After the α-RuCl$_3$ is successfully picked up, we can use more standard parameters to subsequently pick up other 2D materials, e.g. graphene. Using this approach, α-RuCl$_3$ flakes and single-layer graphene were sequentially lifted from an SiO$_2$/Si substrate using a poly(bisphenol A carbonate) (PC) coated glass transfer slide. The PC together with the stack were flipped onto an Si/SiO$_2$ (285 nm Si) substrate held at 150 °C. Indium alloy contacts were placed on the graphene using a micro soldering technique[29] to provide electrical contacts for STM measurements. This technique preserves sample quality compared to lithography methods. See Fig. S1 for diagrammatic procedure.

***Scanning Tunneling Microscopy and Spectroscopy***: All STM/STS measurements were carried out on a commercial RHK system under ultra-high vacuum conditions. An etched Tungsten tip was prepared and calibrated on a Au(111) single crystal. The topographic images were collected in constant current and bias mode using a feedback loop. The STS point spectra were obtained at constant height under open feedback loop conditions with a modulating bias of 25 mV using a lock-in amplifier. $dI/dV$ maps were extracted from a grid of individual point spectra collected in the vicinity of nanobubbles. All measurements were performed at room temperature to permit direct tunneling into $\alpha$-RuCl$_3$ (which is otherwise too resistive at cryogenic temperatures to permit local tunneling measurements).

***Scanning Near-field Optical Microscopy***: All $s$-SNOM measurements were conducted using a commercial Neaspec system under ambient conditions using commercial Arrow$^{TM}$ AFM probes with a nominal resonant frequency of $f = 75$ kHz. Three tunable continuous wave quantum cascade lasers produced by Daylight Solutions were used, collectively spanning wavelengths from 4 to 11 $\mu$m. The detected signal was demodulated at the third harmonic of the tapping frequency in order to minimize background contributions to the scattered light. Simultaneous measurements of the scattering amplitude and phase were performed through use of a pseudoheterodyne interferometer.

***Ab-initio Calculations of graphene/$\alpha$-RuCl$_3$ Heterostructures***: The *ab initio* calculations were performed within the Vienna Ab initio Simulation Package (VASP)[30] using a projector-augmented wave (PAW) pseudopotential in conjunction with the Perdew–Burke–Ernzerhof (PBE)[31] functionals and plane-wave basis set with energy cutoff at 400 eV. For the heterostructures with graphene and monolayer $\alpha$-RuCl$_3$, we used a hexagonal supercell containing 82 atoms (composed of a $5 \times 5$ graphene supercell and $\sqrt{3} \times \sqrt{3}$ $\alpha$-RuCl$_3$ supercell).

The resulting strain is ~2.5% for the α-RuCl$_3$ monolayer. The surface Brillouin zone was sampled by a $3 \times 3 \times 1$ Monkhorst–Pack k-mesh. A vacuum region of 15 Å was applied to avoid artificial interaction between the periodic images along the z direction. Because of the absence of strong chemical bonding between layers, van der Waals density functional in the opt88 form[32] was employed for structural optimization. All structures were fully relaxed until the force on each atom was less than 0.01 eV Å$^{-1}$. Spin-orbital couplings are included in the electronic calculations.

With small Bader charges of 7.01 e (out of 8 e) per orbital, the Ru-4d states cannot be considered fully localized, and therefore, the use of large values of $U_{4d}$ is understood as an ad hoc fitting parameter without physical basis. Instead, each Chlorine 3p orbital charge is 7.34 e (out of 7 e), indicating the importance to employ correction on both Ru and Cl elements. The Hubbard U terms are computed by employing the generalized Kohn–Sham equations within density functional theory including mean-field interactions, as provided by the Octopus package[33,34] using the ACBN0[35,36] functional together with the local density approximation (LDA) functional describing the semilocal DFT part. We compute *ab initio* the Hubbard U and Hund's J for the 4d orbitals of Ruthenium and 3p orbital of Chlorine. We employ norm-conserving HGH pseudopotentials to get converged effective Hubbard U values (1.96 eV for Ru 4d orbitals and 5.31 eV for Cl 3p orbitals) with spin-orbital couplings.

**Acknowledgements**

Research at Columbia University was supported as part of the Energy Frontier Research Center on Programmable Quantum Materials funded by the U.S. Department of Energy (DOE), Office of Science, Basic Energy Sciences (BES), under Award No DE-SC0019443. J.Z. and A.R. were supported by the European Research Council (ERC-2015-AdG694097), the Cluster of Excellence "Advanced Imaging of Matter" (AIM) EXC 2056 - 390715994, funding by the Deutsche Forschungsgemeinschaft (DFG, German Research Foundation) under RTG 2247, Grupos Consolidados (IT1249-19) and SFB925 "Light induced dynamics and control of correlated quantum systems". J.Z., and A.R. would like to acknowledge Nicolas Tancogne-Dejean and Lede Xian for fruitful discussions and also acknowledge support by the Max Planck Institute-New York City Center for Non-Equilibrium Quantum Phenomena. The Flatiron Institute is a division of the Simons Foundation. J.Z. acknowledges funding received from the European Union Horizon 2020 research and innovation program under Marie *Skłodowska*-Curie Grant Agreement 886291 (PeSD-NeSL). STM support was provided by the National Science Foundation via grant DMR-2004691. C.R.-V. acknowledges funding from the European Union Horizon 2020 research and innovation programme under the Marie *Skłodowska*-Curie grant agreement No 844271. D.G.M. acknowledges support from the Gordon and Betty Moore Foundation's EPiQS Initiative, Grant GBMF9069. J.Q.Y. was supported by the U.S. Department of Energy, Office of Science, Basic Energy Sciences, Materials Sciences and Engineering Division. S.E.N. acknowledges support from the U.S. Department of Energy, Office of Science, Basic Energy Sciences, Division of Scientific User Facilities. Work at University of Tennessee was supported by NSF grant No 180896.

**Author Contributions**

S.S. performed the STM/STS measurements. S.S., C.R.-V. and D.J.R. conducted the STS analysis. D.J.R. performed all *s*-SNOM measurements and analysis. A.S.M derived analytical forms for the near-field scattering amplitude and simulated near-field images. F.L.R. modelled the near-field data. J.Z. and A.R. performed all DFT calculations and analyzed the results. B.S.J. fabricated the devices and developed the dry stacking procedure with α-RuCl$_3$. J.C.H. and C.R.D. advised device fabrication efforts. M.C., S.E.N., J.Q.Y. and D.G.M. performed growth and characterization of α-RuCl$_3$ single crystals.

## Competing Interests

The authors declare no competing financial interests.

## Data Availability

All data presented in the manuscript are available upon request.

## Supporting Information Available

Supporting Information contains additional details about sample fabrication, STM and AFM topography, auxiliary STS and *s*-SNOM data, and derivations for models of the near-field data.

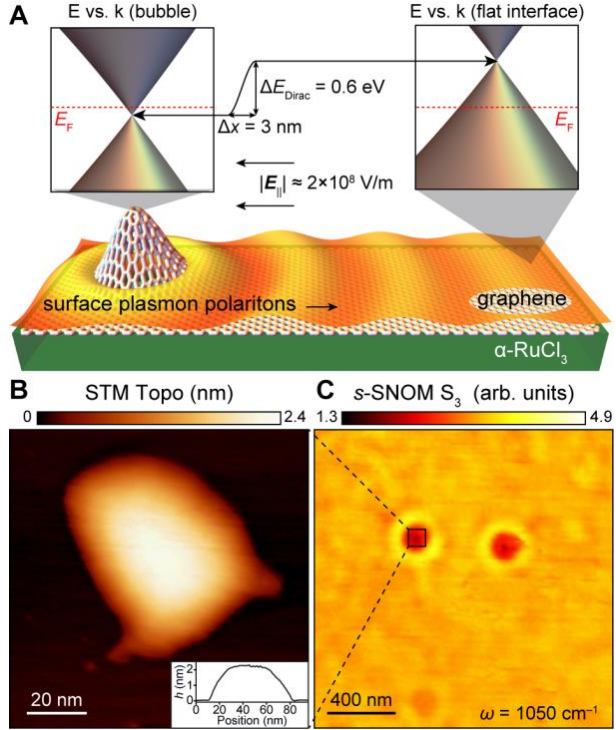

**Figure 1. Overview of joint STM/*s*-SNOM investigation of nanobubbles in graphene/α-RuCl₃ heterostructures.** (**A**) Schematic of Dirac-point energy shift between nanobubbles and clean flat interfaces in graphene/α-RuCl₃ heterostructures. The ~0.6 eV energy shift takes place over a lateral length scale of < 3 nm at the boundary of nanobubbles, generating effective lateral fields of $E_{\parallel} \approx 2 \times 10^8$ V/m (0.2 V/nm). Since the pristine graphene suspended in the nanobubble is intrinsically n-doped, a p-n junction is created at the nanobubble boundary. The associated jump in the graphene conductivity at the perimeter of nanobubble acts as a hard boundary for reflection of surface plasmon polaritons. (**B**) Characteristic STM topographic image of a nanobubble ($V_S = 0.7$ V, $I_t = 50$ pA). The inset shows the one-dimensional cross section of the nanobubble topography. (**C**) Characteristic *s*-SNOM image of two nanobubbles shows circular fringe patterns corresponding to radially-propagating surface plasmon polaritons ($\omega = 990$ cm⁻¹).

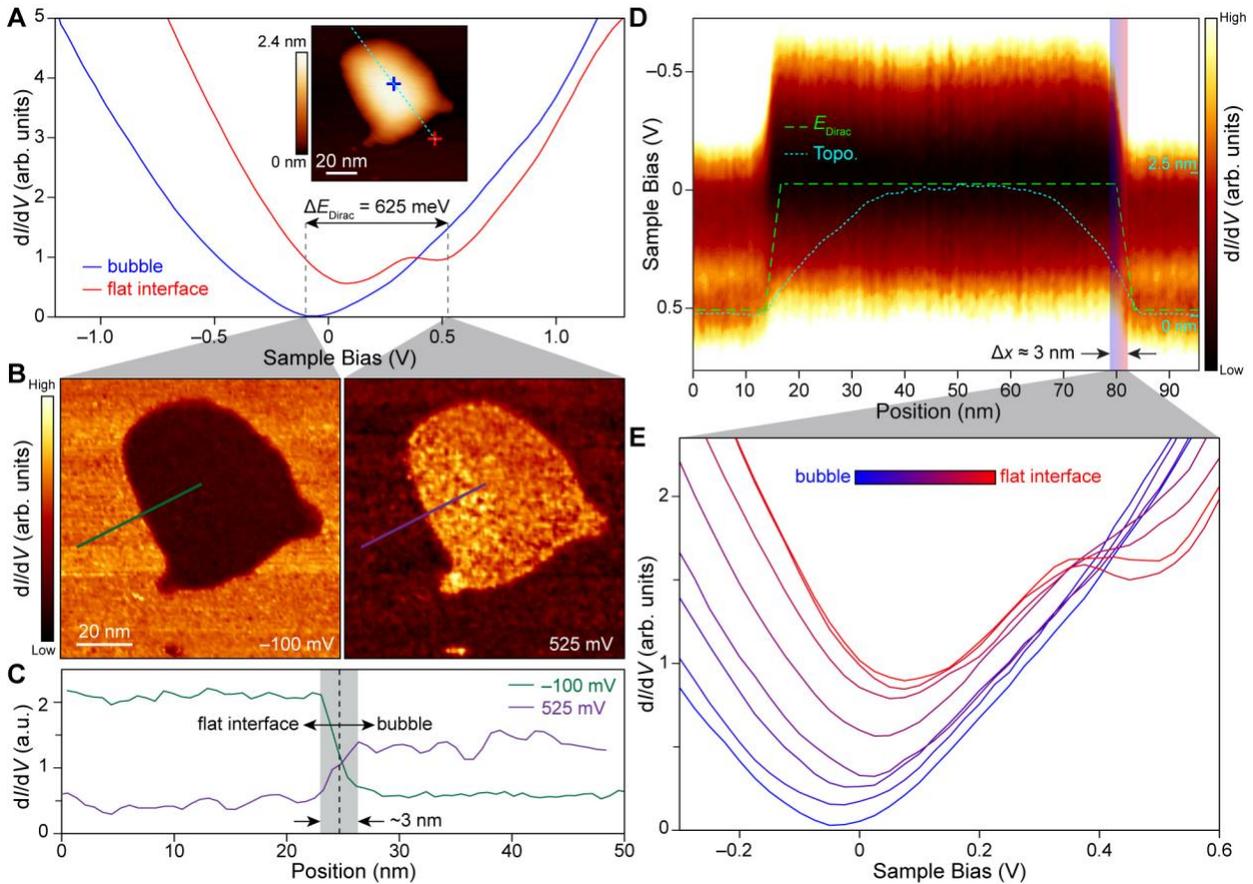

**Figure 2. Electronic structure characterization of nanobubbles in graphene/α-RuCl₃ using STM and STS.** (**A**) Inset: STM topographic image of a graphene nanobubble ($V_S$ = 0.7 V, $I_t$ = 50 pA). Representative d$I$/d$V$ point spectroscopy collected over nanobubbles (blue) and flat graphene/α-RuCl₃ interfaces (red) as indicated by the crosshairs in the inset. Between these two spectra, the graphene Dirac point shifts by 625 meV. (**B**) d$I$/d$V$ maps of a graphene nanobubble conducted at the indicated biases corresponding to the Dirac point energies on the nanobubble (left panel) and the flat interface (right panel) ($V_{AC}$ = 25 mV, $I_t$ = 50 pA). A suppressed LDOS is observed at those biases associated with the local Dirac point energy. (**C**) Linecuts of the d$I$/d$V$ maps shown in (B) following the green and purple lines indicated on the –100 mV and 525 mV maps, respectively. In both instances, the change in the LDOS at the bubble boundary (indicated by the black dashed line) takes place over a lateral length of approximately 3 nm. (**D**) Position-dependent d$I$/d$V$ point spectroscopy collected along the cyan trajectory shown in the inset in (A). The shift in the Dirac point energy occurs over a lateral length scale of ~3 nm as indicated by the region highlighted in partially transparent red and blue. The position-dependence of the Dirac point energy (green dashed line) is superimposed on the topographic line cut (cyan dotted line) showing that the prior has a much more abrupt spatial dependence. (**E**) Sample d$I$/d$V$ point spectra collected at the threshold of a graphene nanobubble corresponding to the red and blue highlighted region in (C).

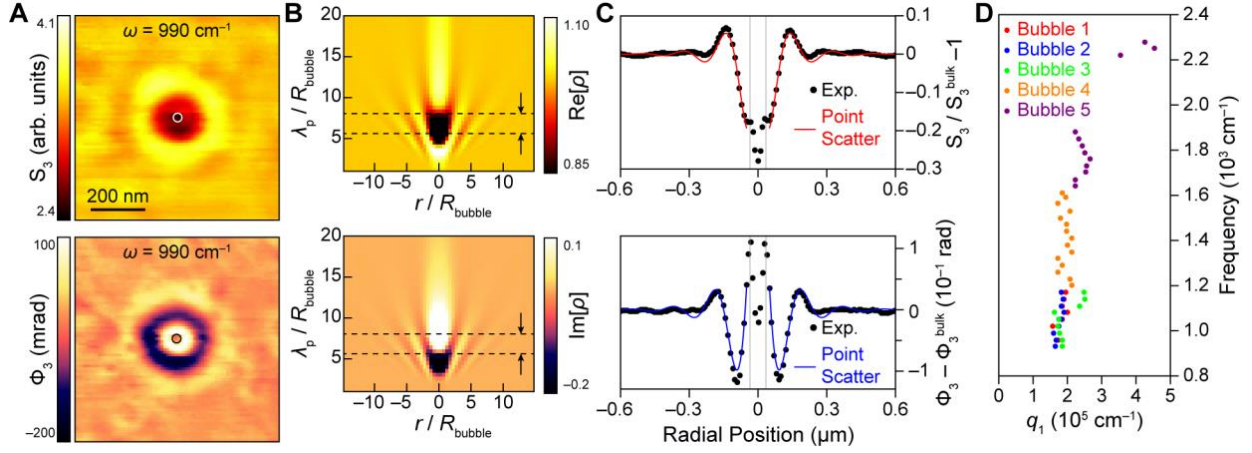

**Figure 3. Characterization of plasmonic response of nanobubbles using *s*-SNOM.** (**A**) *s*-SNOM $S_3$ amplitude (top panel) and $\Phi_3$ phase (bottom panel) collected over a graphene nanobubble ($\omega = 990$ cm$^{-1}$). The bubble perimeter is indicated in each image with a white and black circle, respectively. A characteristic fringe pattern is observed in both the near-field amplitude and phase emanating radially from the bubble. (**B**) Simulated near-field amplitude (top panel) and phase (bottom panel) based on a raster-scanned dipole over a defect with fixed radius $R_{\text{bubble}}$ and a variable SPP wavelength $\lambda_{\text{p}}$. The radial dependence $r/R_{\text{bubble}}$ of both amplitude and phase are shown. The black arrows and black dashed box enclose the regime of $\lambda_{\text{p}}/R_{\text{bubble}}$ that resembles the experimental data. (**C**) Radial line cuts of the images shown in (A) averaged over half-annuli with thicknesses of $\Delta r = 10$ nm. The gray vertical lines indicate the boundaries of the nanobubble. Based on a model that treats the nanobubble as a point scatterer, the radial dependence of the near-field amplitude and phase is simultaneously fit to the real and imaginary components of $-\boldsymbol{A}\left[H_1^{(1)}(\boldsymbol{q}_p r)\right]^2$, respectively ($H_1^{(1)}$ is the Hankel function of first kind of order one, $\boldsymbol{q}_p$ is the complex SPP wavevector, $r$ is the radial coordinate and $\boldsymbol{A}$ is a complex amplitude). (**D**) The corresponding dispersion of SPPs emanating from five different nanobubbles is extracted using the fitting procedure described in (C).

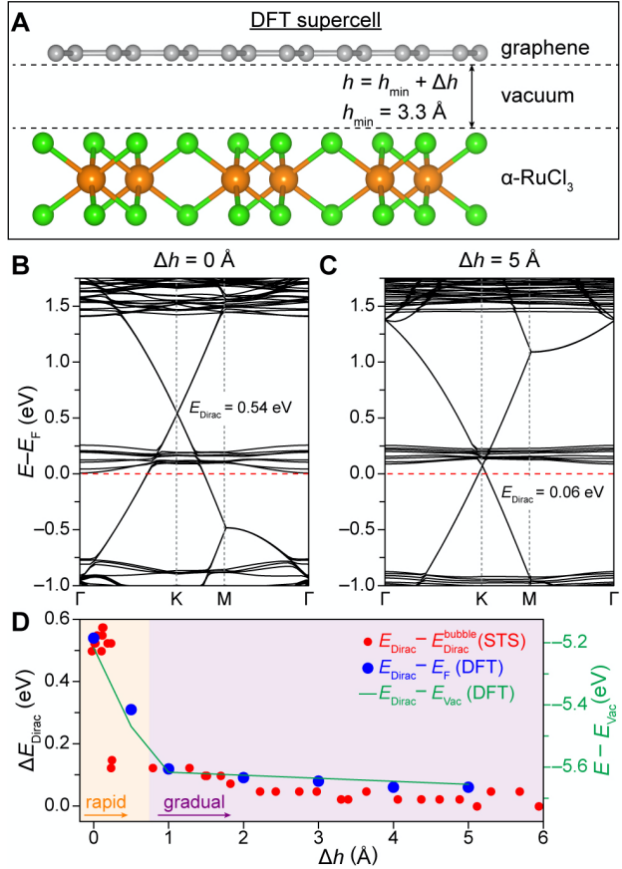

**Figure 4. DFT and STM analysis of interlayer charge transfer in graphene/α-RuCl₃ heterostructures.** (**A**) Side-view of the graphene/α-RuCl₃ heterostructure used in DFT calculations. An equilibrium interlayer separation of $h_{min} = 3.3$ Å is used to model the so-called flat interface observed experimentally. To model the charge transfer behavior between graphene and α-RuCl₃ at the edge of nanobubbles (where the interlayer separation increases gradually), additional calculations are performed using interlayer separations of $\Delta h = h - h_{min} = 0.5, 1, 2, 3,$ 4 and 5 Å. Orange, green and grey spheres indicate Ru, Cl and C atoms, respectively. (**B**) Left panel: DFT-calculated band structure for a graphene/α-RuCl₃ heterostructure with maximal charge transfer (i.e. $h = h_{min} = 3.3$ Å). (**C**) Right panel: Band structure for graphene/α-RuCl₃ heterostructure with $h = h_{min} + 5$ Å, showing minimal interlayer charge transfer. The Fermi levels are set to zero in (B) and (C). (**D**) The shift in $E_{Dirac}$ as a function of interlayer separation is plotted for both experimental (red dots) and theoretical (blue dots) data. The shift in $E_{Dirac}$ relative to the vacuum energy $E_{Vac}$ is plotted in green. The rapid decay is highlighted in orange, while the subsequent gradual decay is highlighted in purple.

# Supplementary Information

**Table of Contents:**

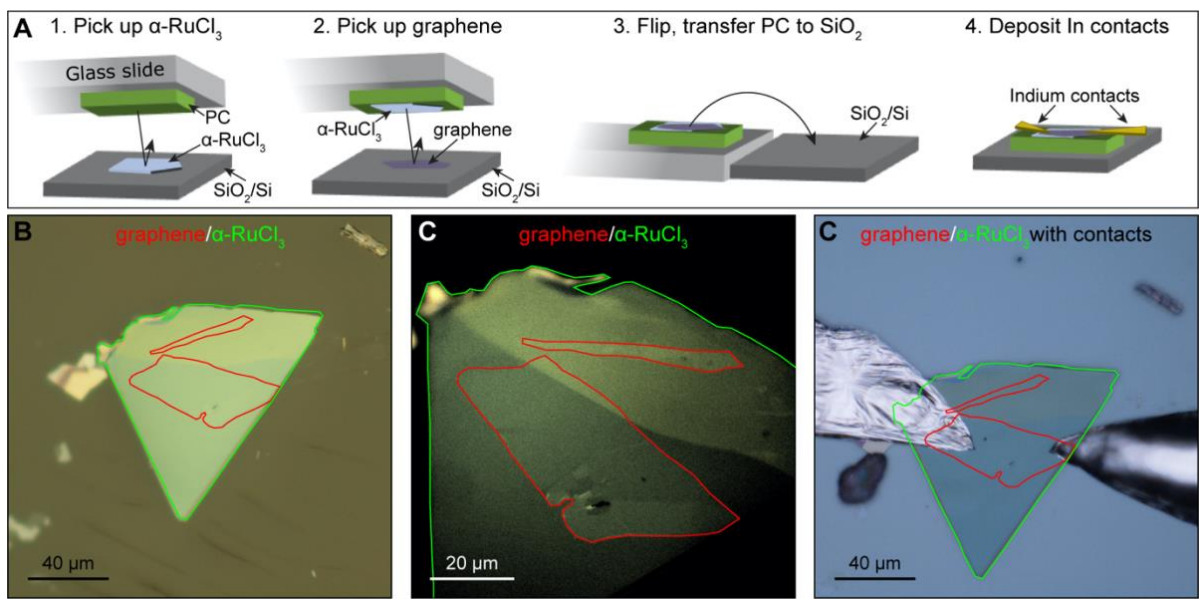

**Figure S1. Graphene/α-RuCl₃ device fabrication. (A)** Diagram of four steps for graphene/α-RuCl₃ device assembly. In the first step, a PC-coated glass slide is used to pick up exfoliated α-RuCl₃ on an SiO₂/Si substrate. In the second step, the α-RuCl₃/PC transfer slide is used to pick up exfoliated graphene. In the third step, the transfer slide is flipped over and the PC is delaminated from the glass slide and placed on a SiO₂/Si chip. In the final step, indium contacts are deposited on the device using a micro soldering approach.[1] **(B)** Optical image of graphene/α-RuCl₃ device with the graphene outlined in red and the α-RuCl₃ outlined in green. **(C)** High contrast magnified image of the stack shown in (C). **(D)** Optical image of the graphene/α-RuCl₃ device after the deposition of indium contacts.

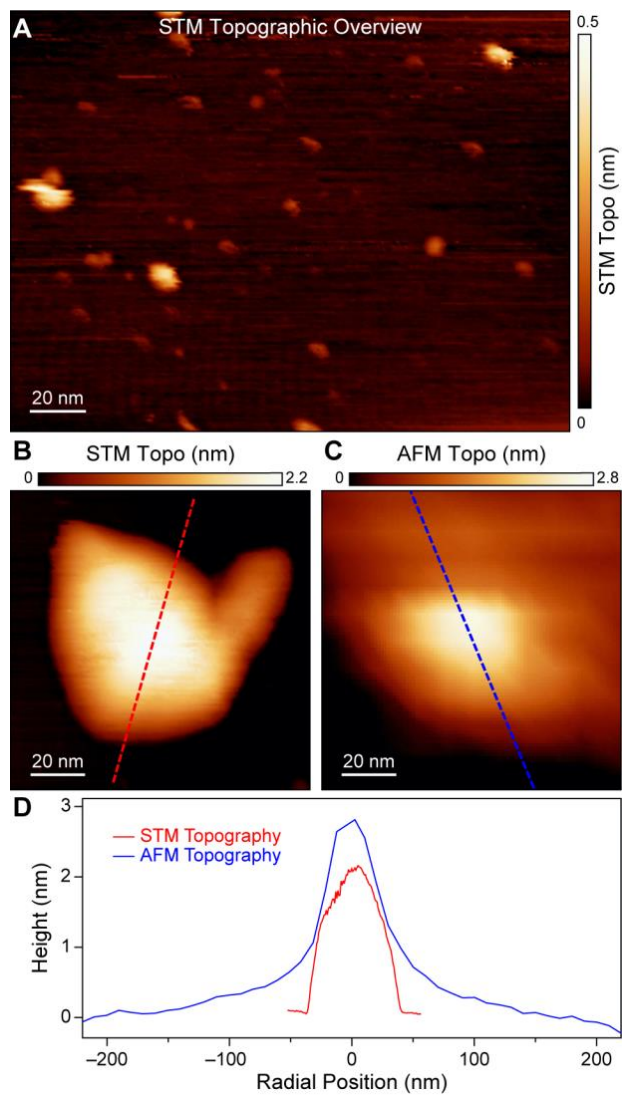

**Figure S2. STM and AFM topographic data. (A)** STM topographic overview of graphene/α-RuCl$_3$ ($V_S = 0.7$ V, $I_t = 50$ pA) showing both flat regions and nanobubbles are present. **(B)** High magnification STM topography of typical graphene nanobubble ($V_S = 0.7$ V, $I_t = 50$ pA). **(C)** High magnification AFM topographic image of typical graphene nanobubble. **(D)** Line profiles based on the topographic images shown in (C) and (D) showing that typical nanobubbles measured in STM have a similar profile to those explored with AFM and $s$-SNOM.

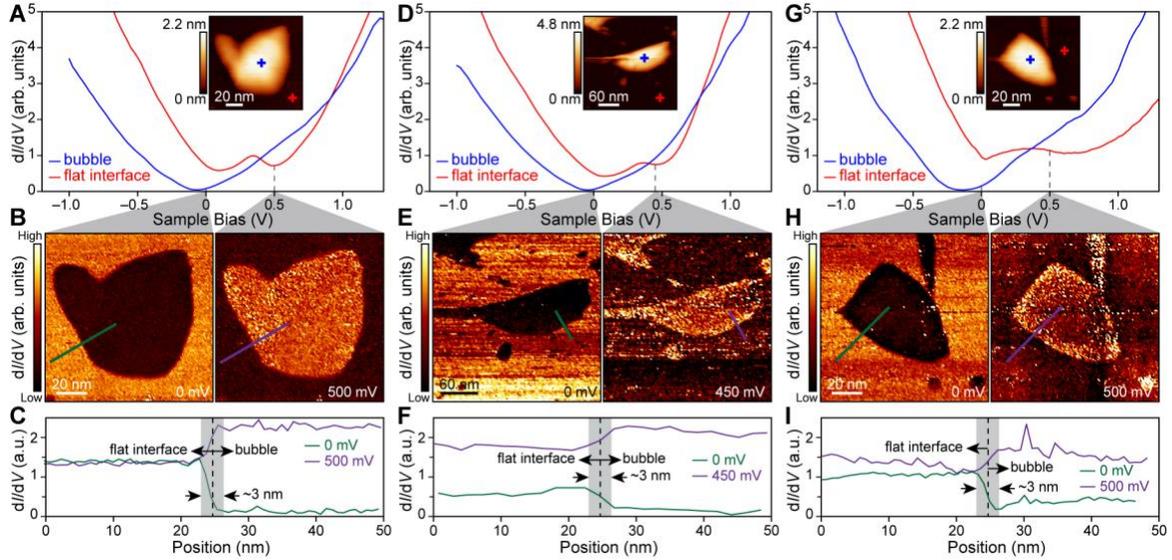

**Figure S3. STM and STS of multiple nanobubbles. (A)** Inset: STM topographic image of a second graphene nanobubble ($V_S = 0.7$ V, $I_t = 50$ pA). Representative d$I$/d$V$ point spectroscopy collected over nanobubbles (blue) and flat graphene/$\alpha$-RuCl$_3$ interfaces (red) as indicated by the crosshairs in the inset. **(B)** d$I$/d$V$ maps of a graphene nanobubble conducted at the indicated biases corresponding to the Dirac point energies on the nanobubble (left panel) and the flat interface (right panel) ($V_{AC} = 25$ mV, $I_t = 50$ pA). A suppressed LDOS is observed at those biases associated with the local Dirac point energy. **(C)** Linecuts of the d$I$/d$V$ maps shown in (B) following the green and purple lines indicated on the –100 mV and 525 mV maps, respectively. In both instances, the change in the LDOS at the bubble boundary (indicated by the black dashed line) takes place over a lateral length of approximately 3 nm. **(D)**, **(E)**, and **(F)** same as (A), (B), and (C) for third graphene nanobubble. **(G)**, **(H)**, and **(I)** same as (A), (B), and (C) for fourth graphene nanobubble.

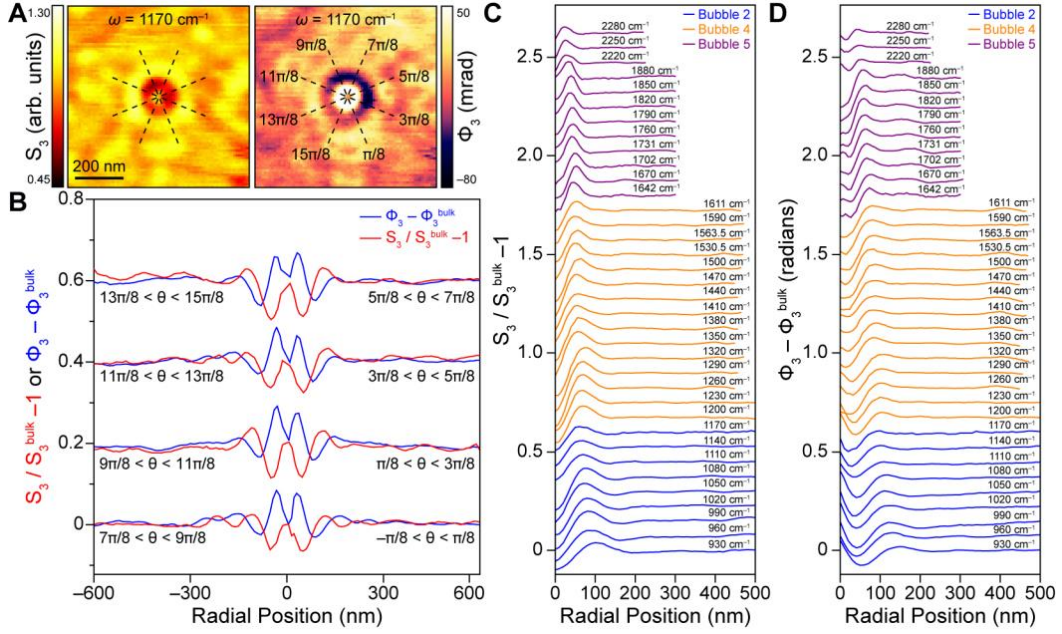

**Figure S4. *s*-SNOM on multiple nanobubbles with *ω*- and angle-dependent near-field linecuts. (A)** *s*-SNOM $S_3$ amplitude (left panel) and $\Phi_3$ phase (right panel) collected over a graphene nanobubble ($\omega = 1170$ cm$^{-1}$). The black dashed lines separate the *s*-SNOM maps into eight angular slices used for the analysis in (B). **(B)** The radial dependence of the *s*-SNOM $S_3$ amplitude (red line) and $\Phi_3$ phase (blue line) integrated over the indicated angles designated in (A). The lack of a systematic angular dependence suggests that $\lambda_p$ fringes do not contribute significantly to the plasmonic response of nanobubbles. **(C)** The radial dependence of the $S_3$ amplitude is shown for frequencies spanning $\omega = 930$ cm$^{-1}$ – $2280$ cm$^{-1}$ collected on bubble 2 (blue lines), bubble 4 (orange lines) and bubble 5 (purple lines) referenced in Fig. 3 of the main manuscript. Since bubbles 1, 2, and 3 all overlap in frequency, only bubble 2 is shown for clarity. All line profiles are truncated at the boundary of the associated nanobubble. **(D)** Same as (C) but for the radial dependence of the $\Phi_3$ phase.

**Supplementary Discussion**

Modeling near-field signal from plasmon reflection at a finite-sized bubble defect

As attested by our experimental results, we model the plasmonic response of a single nanobubble in the graphene/$\alpha$-RuCl$_3$ heterostructure by a local perturbation of the graphene sheet conductivity $\sigma$ with respect to its asymptotic value $\sigma(\infty)$ arising from charge transfer from the $\alpha$-RuCl$_3$ underlayer. We denote the relative inhomogeneity in conductivity due the nanobubble as $\bar{\sigma}(\mathbf{r}) = \sigma(\mathbf{r})/\sigma(\infty)$. To model the position-dependent near-field signal associated with reflections of plasmon polaritons from the defect, we considered the integro-differential equation for the scalar potential $\phi_s$ generated in response to the incident potential $\phi_{\text{probe}}$ of a near-field probe[2]:

$$\left[1 + \frac{1}{2\pi q_s} V * \nabla \cdot \bar{\sigma}(\mathbf{r}) \nabla\right] \phi(\mathbf{r}) = \phi_{\text{probe}}(\mathbf{r}), \quad \phi = \phi_{\text{probe}} + \phi_s. \tag{S1}$$

Here $q_s = i\omega/(2\pi\sigma(\infty))$ parameterizes the asymptotic conductivity away from the defect through its associated plasmon polariton momentum, $V(r) = 1/(\kappa\, r)$ is the Coulomb kernel screened by permittivity $\varepsilon$ of the proximate $\alpha$-RuCl$_3$ underlayer with $\kappa = (\varepsilon + 1)/2$, and the asterisk ($*$) denotes the spatial convolution over the in-plane coordinate $\mathbf{r} = (x, y)$. As an example, we choose $\bar{\sigma}(\mathbf{r}) \equiv 1 + \delta\Lambda(r/R_{\text{bubble}})$, where $\delta$ is the characteristic magnitude of the conductivity fluctuation at the nanobubble, $R_{\text{bubble}}$ is its width, and $\Lambda(r) = 1 - \theta(r - 1)$ is taken as a step function of unit radius and height. We solved Eq. (S1) through expansion in an orthonormal basis of plane waves $\phi_j = A_j e^{i\mathbf{q}_j \cdot \mathbf{r}}$ periodic in a 2D square cell $x, y \in [-L/2, L/2]$, with $A_j$ a normalization constant and $L \gg R_{\text{bubble}}$. If we assemble the Fourier momenta $\mathbf{q}_j$ and the Fourier coefficients $\tilde{\phi}_j = \langle \phi_j | \phi \rangle \equiv \int \phi_j^*(\mathbf{r})\phi(\mathbf{r})d^2r$ into column vectors $\vec{q}$ and $\vec{\phi}$, respectively, then $\langle \phi_i | V * | \phi_j \rangle = 2\pi/(\kappa q_i)\,\delta_{ij}$ with $\delta_{ij}$ the Kronecker delta, and these vectors must obey the equation

$$\vec{\phi} = \left[q_s^* - \left(\delta\vec{Q} + \text{diag}\,|\vec{q}|\right)\right]^{-1} q_s^*\, \vec{\phi}_{\text{probe}}, \tag{S2}$$

where $q_s^* = \kappa\, q_s$ defines the screened polariton momentum, and $\vec{Q}$ is the scattering matrix with the elements

$$Q_{ij} = \left(\hat{\mathbf{q}}_i \cdot \mathbf{q}_j\right)\left\langle \phi_i \left| \Lambda\left(\frac{r}{R_{\text{bubble}}}\right) \right| \phi_j \right\rangle. \tag{S3}$$

We defined another matrix-valued function $\vec{G}$ by $\vec{\phi}_s = \vec{G}\,\vec{\phi}_{\text{probe}}$. From Eq. (S2), we obtain

$$G_{ij} = \left\langle \phi_i \left| \left[q_s^* - \left(\delta\vec{Q} + \text{diag}\,|\vec{q}|\right)\right]^{-1}\left(\delta\vec{Q} + \text{diag}\,|\vec{q}|\right) \right| \phi_j \right\rangle. \tag{S4}$$

For a translationally invariant system, $\delta = 0$, where the momentum is conserved, only the diagonal matrix elements are nonzero. They can be understood as "in-plane" reflection coefficients, and are related to the conventional Fresnel coefficients $r_P(\omega, q)$ by $-G_{jj} =$ $r_P(\omega, q = |\mathbf{q}_j|)$. Therefore, $\mathrm{Im}\,(-G_{jj}) = f(\omega, \mathbf{q}_j)$ has maxima at the same plasmon polariton momenta $|\mathbf{q}_j| = \mathrm{Re}\,q_s^*$ as $\mathrm{Im}\,r_P$. However, our interest concerns $\delta \neq 0$.

Previous work[3] has established a leading order approximation to the complex-valued near-field signal $\rho$ scattered by a probe, given by the Fourier integral:

$$\rho \sim -\frac{1}{2\pi}\int d^2q\;|\mathbf{q}|\;\tilde{\phi}_{\mathrm{probe}}(\mathbf{q})\;\tilde{\phi}_s(\mathbf{q}) \tag{S5}$$

where $\tilde{\phi}_{\mathrm{probe}}$ and $\tilde{\phi}_s$ denote Fourier transforms of the respective potentials with respect to in-plane (vector) momenta $\mathbf{q}$ evaluated at the surface plane of the sample. The notation $\rho$ used here for the near-field signal affirms its connection to the so-called photonic density of states as motivated in ref. [3]. In our case where $\mathbf{q}_j$ describe a uniformly spaced grid of momenta spanning the "first Brillouin zone" of the simulation domain, Eq. (S5) is readily evaluated by:

$$\rho \sim \frac{1}{2\pi}\vec{\phi}_{\mathrm{probe}}^{\,T}\,\mathrm{diag}\,|\vec{q}|\;\vec{G}(q_s^*, R_{\mathrm{bubble}})\,\vec{\phi}_{\mathrm{probe}}. \tag{S6}$$

Here we highlight that the dependence on screened plasmon wavevector and nanobubble size resides in $\vec{G}(q_s^*, R_{\mathrm{bubble}})$, which encodes the associated inhomogeneous optical response.

We developed a Python-language computer code implementing the above equations taking advantage of public-domain libraries and we used it to carry out a series of numerical simulations. For simplicity, we approximated $\phi_{\mathrm{probe}}(\mathbf{r})$ by a potential of a point dipole placed a small distance $z_{\mathrm{probe}}$ away from graphene[4]. Given an in-plane probe position $\mathbf{r}_{\mathrm{probe}}$, the relative strength $\delta$ of the perturbation due to the nanobubble, and the nanobubble radius $R_{\mathrm{bubble}}$, the code computes the complex-valued amplitude and phase of $\rho$. We take $z_{\mathrm{probe}} \approx a \approx 30$ nm to appropriately treat the incident field from the near-field probe with apex radius $a$. Informed by our STS results demonstrating near uniform suppression (on the scale of both $a$ and the unperturbed polariton wavelength) of the graphene Fermi level to near the Dirac point across the entire nanobubble, we take $\delta \approx -1$ to denote complete suppression of free carrier conductivity. Meanwhile, as a representative case, we select $R_{\mathrm{bubble}} = 30$ nm $\approx a$. Results presented in Fig. 3B of the main text were obtained by computing $\rho$ for numerous values of $q_s^*$ and the probe position $r \equiv |\mathbf{r}_{\mathrm{probe}}|$, and normalizing the result by its value at $\rho(r \to \infty)$, thus highlighting contrasts due solely to the nanobubble-scattered field. The result can be straightforwardly understood as uniquely a function of three dimensionless ratios, $z_{\mathrm{probe}}/R_{\mathrm{bubble}}$, $\lambda_p/R_{\mathrm{bubble}}$, and $r/R_{\mathrm{bubble}}$, where $\lambda_p \equiv 2\pi/q_s^*$ defines the wavelength of the plasmon polariton in the bulk of graphene. The select results shown in Fig. 3B of the main text are broadly characteristic of the case where $z_{\mathrm{probe}} \sim R_{\mathrm{bubble}}$, and are therefore well representative of the infrared nano-imaging results for nanobubbles characterized in this work.

Derivation of scattering amplitude for plasmonic point-scatterer

In this section we utilize notations common to the previous section, where possible. The polariton scattering problem Eq. (S1) admits an analytic solution for the total field $\phi = \phi_{\mathrm{probe}} + \phi_s$ in the case that the excitation field $\phi_{\mathrm{probe}}$ and the "defect" in graphene optical conductivity $\Delta\bar{\sigma}(\mathbf{r}) = \bar{\sigma}(\mathbf{r}) - 1$ take the form of a point source and a point scatterer, respectively. Provided that the defect and source are "not too strong", a perturbation theory can be applied. The condition for its self-consistency will be discussed in context of the result. In this case, it is convenient to rewrite Eq. (S1) in an operator notation:

$$\left[1 - (\hat{L}_0 + \epsilon \cdot \hat{L}')\right]\phi = \phi_{\text{probe}}$$

$$\text{where} \quad \hat{L}_0 \equiv -\frac{1}{2\pi q_s} V * \nabla^2 \quad \text{and} \quad \hat{L}' \equiv -\frac{1}{2\pi q_s} V * \nabla \cdot \frac{1}{\epsilon} \Lambda(\mathbf{r} - \mathbf{r}_0)\nabla. \tag{S7}$$

Here $\Lambda(\mathbf{r}) \approx A_s \, \delta(\mathbf{r})$ denotes the profile selected to describe the defect centered at lateral coordinate $\mathbf{r}_0$, with $A_s$ its integral weight, in units of area, and $\delta(\mathbf{r})$ a Dirac delta function. Meanwhile, taking $\epsilon \ll 1$ supplies a perturbation expansion provided that $\hat{L}'\phi_{\text{probe}}$ remains "small":

$$\phi = \left[1 - (\hat{L}_0 + \epsilon \cdot \hat{L}')\right]^{-1}\phi_{\text{probe}}$$
$$\approx \left[\hat{G}_0 + \epsilon \hat{G}_0 \hat{L}' \hat{G}_0 + O(\epsilon^2)\right]\phi_{\text{probe}},$$
$$\text{with} \quad \hat{G}_0 \equiv \left(1 - \hat{L}_0\right)^{-1}. \tag{S8}$$

Here $\hat{G}_0$ defines a "bare" propagator for plasmon polaritons. This propagator can be obtained through a Fourier representation of Eq. (S7) with respect to the in-plane wavevector $\mathbf{q}$, whereby:

$$\phi(\mathbf{r}) = \int \frac{d^2q}{2\pi} e^{i\mathbf{q}\cdot\mathbf{r}} \phi(\mathbf{q}), \quad \phi(\mathbf{r}) = \int \frac{d^2q}{2\pi} e^{i\mathbf{q}\cdot\mathbf{r}} \phi(\mathbf{q}), \quad \text{and} \quad \hat{L}_0 = |\mathbf{q}|/q_s^*. \tag{S9}$$

Here we use the unitary Fourier transform. In the Fourier domain, the propagator is naively then expressed by $G_0(q) = q_s^*/(q_s^* - q)$. However, note that this form of the propagator $\hat{G}_0\phi_{\text{probe}}$ can only generate the inhomogeneous part of solutions $\phi$, to which any arbitrary homogeneous part $\phi_h$ for which $(1 - \hat{L}_0)\phi_h = 0$ can also be added, e.g. $\phi = \hat{G}_0\phi_{\text{probe}} + \phi_h$, however necessary to satisfy the prescribed boundary conditions. For the case of an open system of graphene on $\alpha$-RuCl$_3$ illuminated by a localized probe, we will demand an outgoing radiation condition for $\phi$. In other words, $\phi$ must vanish at infinite distance, and (polariton) waves must propagate outwards, with complex phase decreasing uniformly with distance from the source. To this end, we can augment the propagator as follows to enforce this condition. We first consider a point source placed at the origin, $\phi_{\text{probe}}(\mathbf{r}) = A_p \delta(\mathbf{r})$, where $A_p$ denotes the integral weight of the excitation (in units of area), for which $\phi_{\text{probe}}(q) = A_p/2\pi$. The inhomogeneous part of the solution is given by:

$$\left[\hat{G}_0\phi_{\text{probe}}\right](\mathbf{r}) = \frac{A_p}{2\pi} \int \frac{d^2q}{2\pi} e^{i\mathbf{q}\cdot\mathbf{r}} G(q)$$
$$= \frac{A_p}{(2\pi)^2} \int_0^\infty dq \, q \left(\frac{q_s^*}{q_s^* - q}\right) \int_0^{2\pi} d\theta \, e^{iqr\cos\theta}$$

$$= -\frac{A_p q_s^*}{2\pi} \int_0^\infty dq \, \frac{q}{q - q_s^*} J_0(qr) \tag{S10}$$

$$= -\frac{A_p q_s^{*2}}{2\pi} \left[ \frac{1}{q_s^* r} - \frac{\pi}{2} \left( Y_0(q_s^* r) + \mathbf{H}_0(q_s^* r) \right) \right]$$

Here we have applied identity (2.12.3.11) of ref. [5] to the case of the Bessel function of order $\nu = 0$, where $J_0(\dots)$, $Y_0(\dots)$ and $\mathbf{H}_0(\dots)$ denote Bessel functions of the first and second kinds and the Struve-H function, respectively, all of order $\nu = 0$. For distances $r_0 \gg \lambda_p = 2\pi/Re[q_s^*]$, the sum in brackets is very nearly equal to $-\pi Y_0(q_s^* r)$, which can be identified as the inhomogeneous part of the solution to the wave equation with open boundary conditions. The outgoing wave condition is therefore enforceable by an added homogeneous part $\phi_h \propto i\pi J_0(q_s^* r)$, in which case the term in square brackets becomes very nearly equal to $i\pi H_0^1(q_s^* r)$, with $H_0^1(\dots)$ the Hankel function of the first kind of order $\nu = 0$, representing an outgoing cylindrical wave. Consequently, we forthwith augment the Fourier space propagator to enforce our prescribed boundary conditions:

$$G(q) = q_s^* \left( \frac{1}{q_s^* - q} + i\pi \delta(q - q_s^*) \right),$$

$$\text{so that} \quad G_0(r) \approx \frac{i}{2} q_s^{*2} H_0^1(q_s^* r). \tag{S11}$$

Here the Dirac delta function supplies the homogeneous component in Fourier space. Deviations not captured by this functional form at distances $r \to 0$ associate with the "local" metallic response of the plasmonic medium, which supply screening of the incident divergent field as $q_s^* \to 0$ in the limit where surface conductivity diverges to infinity. While this physical behavior is not captured by a mere wave solution, it remains inessential to our experimental results.

Meanwhile, the Fourier space representation for $\hat{L}'$ operating on a function $f(\mathbf{q})$ is:

$$[\hat{L}' f](\mathbf{q}) = \frac{1}{\epsilon q_s^*} \hat{\mathbf{q}} \cdot \int d^2 q' \, 2\pi \, \Lambda(\mathbf{q} - \mathbf{q}') \, \mathbf{q}' \, f(\mathbf{q}') \tag{S12}$$

Here $\hat{\mathbf{q}}$ denotes a unit wavevector, and real-space multiplication by $\Lambda(\mathbf{r} - \mathbf{r}_0)$ within $\hat{L}'$ is transformed by the convolution theorem into an integral kernel $2\pi \, \Lambda(\mathbf{q} - \mathbf{q}')$. Next, we apply the Fourier representation of the defect profile $\Lambda(q) = e^{-i\mathbf{q} \cdot \mathbf{r}_0}/2\pi$ representing the Dirac delta function centered at $\mathbf{r}_0$, obtaining:

$$[\hat{L}' f](\mathbf{q}) = \frac{A_s}{\epsilon q_s^*} \hat{\mathbf{q}} \cdot \int d^2 q' \, e^{-i(\mathbf{q} - \mathbf{q}') \cdot \mathbf{r}_0} \, \mathbf{q}' \, f(q')$$

$$= \frac{A_s}{\epsilon q_s^*} e^{-i\mathbf{q} \cdot \mathbf{r}_0} \int_0^\infty dq' \, q'^2 f(q') \int_0^{2\pi} d\theta' \cos(\theta' - \theta) \, e^{iq' r_0 \cos\theta'}$$

$$= \frac{2\pi i \, A_s}{\epsilon q_s^*} \cos\theta \, e^{-iq r_0 \cos\theta} \int_0^\infty dq' \, q'^2 \, J_1(q' r_0) \, f(q'). \tag{S13}$$

Here $J_1$ denotes the Bessel function of the first kind of order $\nu = 1$, and $\theta'$ and $\theta$ denote the angles subtended between the position vector $\mathbf{r}_0$ and the incoming and outgoing wavevectors $\mathbf{q}'$ and $\mathbf{q}$, respectively. Here we have also assumed $f(\mathbf{q})$ to be an isotropic function. Since the defect-scattered field is given by $\Delta\phi(\mathbf{r}) = \epsilon \hat{G}_0 \hat{L}' \hat{G}_0\, \phi_{\text{probe}}(\mathbf{r})$, then $f = \hat{G}_0\, \phi_{\text{probe}}$, and our point source at the origin is compatible with this assumption. The latter integral in Eq. (S13) can now be evaluated:

$$f(q) = \left[\hat{G}_0\, \phi_{\text{probe}}\right](q) = q_s^*\left(\frac{1}{q_s^* - q} + i\pi\delta(q - q_s^*)\right)\frac{A_p}{2\pi}, \quad \text{so that}$$

$$\int_0^\infty dq'\, q'^2\, J_1(q' r_0)\, f(q') = \frac{A_p q_s^*}{2\pi}\left[i\pi q_s^{*2}\, J_1(q_s^* r_0) + \int_0^\infty dq'\, \frac{q'^2}{q_s^* - q'} J_1(q' r_0)\right]$$

$$= \frac{A_p q_s^*}{2\pi}\left[i\pi\, J_1(q_s^* r_0) - \frac{\partial}{\partial r_0}\int_0^\infty dq'\, \frac{q'}{q_s^* - q'} J_0(q' r_0)\right]$$

$$= \frac{A_p q_s^{*2}}{2\pi}\left\{i\pi q_s^*\, J_1(q_s^* r_0) - \frac{\partial}{\partial r_0}\left[\frac{1}{q_s^* r} - \frac{\pi}{2}\left(Y_0(q_s^* r) + \mathbf{H}_0(q_s^* r)\right)\right]\right\}. \tag{S14}$$

Noting again that the sum in square brackets is very approximately equal to $-\pi\partial_{r_0} Y_0(q_s^* r) = +\pi q_s^* Y_1(q_s^* r)$, the sum in curled brackets is also very nearly equal to $i\pi q_s^* H_1^1(q_s^* r_0)$, a Hankel function of the first kind of order $\nu = 1$. Inserting this wave function back into Eq. (13), we have:

$$\left[\hat{L}' \hat{G}_0\, \phi_{\text{probe}}\right](\mathbf{q}) = \left(\frac{2\pi i\, A_s}{\epsilon q_s^*}\cos\theta\, e^{-iqr_0\cos\theta}\right) \times \frac{A_p q_s^{*2}}{2\pi} \times i\pi q_s^* H_1^1(q_s^* r_0)$$

$$= -\frac{i\pi}{\epsilon} A_s A_p q_s^{*2} H_1^1(q_s^* r_0)\cos\theta\, e^{-iqr_0\cos\theta} \tag{S15}$$

The field scattered by the defect can now be evaluated at the origin $\mathbf{r} = \mathbf{0}$, coinciding with the location of the probe field, as:

$$\Delta\phi(\mathbf{r} = \mathbf{0}) = \epsilon \int \frac{d^2 q}{2\pi}\, \left[\hat{G}_0 \hat{L}' \hat{G}_0\, \phi_{\text{probe}}\right](\mathbf{q})$$

$$= -i\pi A_s A_p q_s^{*3} H_1^1(q_s^* r_0)\int_0^\infty dq\, q\, \left(\frac{1}{q_s^* - q} + i\pi\delta(q - q_s^*)\right)\int_0^{2\pi}\frac{d\theta}{2\pi}\cos\theta\, e^{-iqr_0\cos\theta}$$

$$= +i\pi A_s A_p q_s^{*3} H_1^1(q_s^* r_0)\left[i\pi q_s^*\, J_1(q_s^* r_0) + \int_0^\infty dq\, \frac{q}{q_s^* - q} J_1(qr_0)\right] \tag{S16}$$

$$\approx i\pi A_s A_p q_s^{*3} H_1^1(q_s^* r_0) \times -\frac{\partial}{\partial r_0} i\pi\, H_0^1(q_s^* r)$$

$$\approx -\left(A_s q_s^{*2}\right)\left(A_p q_s^{*2}\right)\left(\pi H_1^1(q_s^* r_0)\right)^2. \tag{S17}$$

Here we have identified the term in square brackets as proportional to the $r_0$-derivative of our augmented propagator $G_0(r = r_0)$, for which we readily supply the outgoing wave approximation (Eq (11)).

We note that the two leading dimensionless terms in parentheses in Eq. (S17) scale as the perturbation area in comparison to the plasmon wavelength $\lambda_p = 2\pi/q_s^*$ squared. In the context where graphene nanobubbles scatter plasmon polariton fields with momentum $q_s^*$, the defect area is described by $A_s = -\pi R_{\text{bubble}}^2$ (negation implying a deficit of conductivity) and the perturbation treatment applied here is self-consistent so long as $R_{\text{bubble}} \ll \lambda_p$. Since excitation from the near-field probe may be described by $A_p \sim a^2$ with $a$ the probe tip radius, the condition $a < \lambda_p$ implies the perturbation treatment here should be a particularly robust description of our experiments. Our nano-imaging experiments approximately detect the vertically polarized field scattered on the graphene surface. This field is proportional to instantaneous surface charge on the graphene, which is in turn proportional to $\Delta\phi(\mathbf{r})$. Taking $r_0$ as the probe-nanobubble separation distance, we can therefore directly apply the complex-valued functional form $H_1^1(q_s^* r_0)^2$ to fit the line-profiles presented in Fig. 3C of the main text. This form is characterized by alternating fringes with an apparent spatial period of $\lambda_p/2$, owing to round-trip traversal of polariton fields over a cumulative distance $2r_0$ between the probe and the nanobubble and back. This formalism therefore supplies a quantitative means to extract plasmon polariton momentum and wavelength directly from our nano-infrared images.